\documentclass[showpacs,twocolumn,superscriptaddress]{revtex4-2}
\usepackage{doi}
\usepackage{hyperref}
\hypersetup{
colorlinks=true,        
linkcolor=blue,         
citecolor=cyan,         
}

\usepackage{graphicx}
\usepackage{dcolumn}
\usepackage{bm}
\usepackage{color}
\usepackage{enumitem}
\usepackage{amsmath}
\usepackage{amssymb}

\begin{document}
\title{Epicyclic oscillations and accretion disk around a special Buchdahl-inspired spacetime}

\author{Mirzabek Alloqulov}
\email{malloqulov@gmail.com}
\affiliation{School of Physics, Harbin Institute of Technology, Harbin 150001, People’s Republic of China} 
\affiliation{University of Tashkent for Applied Sciences, Str. Gavhar 1, Tashkent 100149, Uzbekistan}

\author{Mubasher Jamil}
\email{mjamil@sns.nust.edu.pk}
\affiliation{School of Natural Sciences, National University of Sciences and Technology, Islamabad 44000, Pakistan}

\author{Sanjar Shaymatov}
\email{sanjar@astrin.uz}
\affiliation{Institute of Fundamental and Applied Research, National Research University TIIAME, Kori Niyoziy 39, Tashkent 100000, Uzbekistan}
\affiliation{Institute for Theoretical Physics and Cosmology, Zhejiang University of Technology, Hangzhou 310023, China}
\affiliation{Western Caspian University, Baku AZ1001, Azerbaijan}

\author{Qiang Wu}
\email{wuq@zjut.edu.cn}
\affiliation{Institute for Theoretical Physics and Cosmology, Zhejiang University of Technology, Hangzhou 310023, China}
\affiliation{United Center for Gravitational Wave Physics (UCGWP), Zhejiang University of Technology, Hangzhou, 310032, China}

\author{Mustapha Azreg-A\"{\i}nou}
\email{azreg@baskent.edu.tr (corresponding author)}
\affiliation{Ba\c{s}kent University, Engineering Faculty, Ba\u{g}l\i ca Campus, 06790-Ankara, Turkey}

\date{\today}
\begin{abstract}

In this paper, we consider the Buchdahl-inspired spacetime metric and investigate its aspects using the quasiperiodic oscillations and the accretion disk due to the accreting matter. First, we focus on analyzing the geodesics of particles around the Buchdahl-inspired spacetime, together with the conserved quantities such as specific energy and angular momentum for massive particles orbiting on the innermost stable circular orbits (ISCOs). We show that the effect of the Buchdahl parameter $\tilde{k}$ increases as the radii of the ISCO orbits decrease, resulting in shifting orbits toward the central object compared to the Schwarzschild black hole case. We also consider astrophysical epicyclic oscillations and derive their general expressions using implications of circular motion of massive particles around the Buchdahl-inspired spacetime. Further, we explore the astrophysical implications of observational higher-frequency QPOs of the selected galactic microquasars of X-ray binary systems to obtain the best-fit constraints on the Buchdahl-inspired spacetime parameters. Finally, we consider the accretion disk around the Buchdahl-inspired spacetime using implications of the ISCO parameters that define the accretion disk's inner edge. We explore radiation properties of the accretion disk and redshifted image and intensity of a lensed accretion disk around the Buchdahl-inspired spacetime.  

\end{abstract}

\maketitle
\footnotesize

\section{Introduction}

Einstein’s theory of general relativity (GR) has been, so far, the most successful theory of gravitation that has passed every experimental or observational test, most notably and recently, the LIGO/Virgo \cite{LIGOScientific:2016aoc} and EHT experiments \cite{EventHorizonTelescope:2019ths,2023arXiv231108680T}. Historically, GR was tested in the weak field gravity regime at the solar system level such as the perihelion shift of Mercury, the gravitational deflection of light by the Sun, also the expansion of the universe, and much later using binary pulsars giving indirect evidence of gravitational waves \cite{Bambi:2015kza,Stairs:2003eg,Ishak:2018his}. Despite its numerous successes, GR is an incomplete theory and has certain limitations due to its inability to explain, most notably, the mysterious nature of dark matter, dark energy, and the spacetime singularities among several others. These problems need to be explored by extending GR via curvature or quantum corrections in the Einstein-Hilbert action by making use of the effective field theory approach, which ultimately would lead to quantum-corrected black holes \cite{Lewandowski:2022zce,Jusufi:2022uhk,Mele:2021hro}.

In the literature, several modified gravity theories have been proposed over the years; most relevant to our purpose is the $f(R)$ theory \cite{Ferraro:2012wp}. We are interested in a particular case where $f(R)\sim R^2$. Gravitational models, which involve the term $R^2$ in the gravitational Lagrangian, are termed \textit{quadratic gravity} and may typically appear in the form $R+R^2+\text{curvature corrections}$, in addition to the matter action \cite{Pravda:2024uyv}. However, ignoring the Einstein-Hilbert action and the curvature corrections yields a \textit{pure} $R^2$ gravity model. This theory is both ghost-free and scale-invariant. In 1961, Buchdahl investigated the pure $R^2$ gravity with vacuum to obtain a solution of the resulting field equations \cite{1962NCim...23..141B,Nguyen:2023qux}, however, his analysis of equations lead to a nonlinear second-order ordinary differential which he left unsolved. Nguyen was able to solve the Buchdahl’s ODE and obtained a novel class of metrics \cite{Nguyen:2022bmj}. Among this wide class, Nugyen also obtained an asymptotically flat metric, named \textit{special} Buchdahl inspired metric, involved a Buchdahl parameter $\tilde k$ and a Schwarzschild radius $r_s$ \cite{Nguyen:2022blj}. For $\tilde k=0$, the metric reduces to the Schwarzschild spacetime. The \textit{special} Buchdahl inspired metric is an interesting metric as it represents different geometries including a Schwarzschild black hole, wormhole, and even a naked singularity for different values of parameter $\tilde k$ \cite{Nguyen:2023kwr}. In the literature, the \textit{special} Buchdahl inspired metric has been tested using S2 star observations and solar system observations \cite{Zhu:2024oxz}. An axisymmetric generalization of the special Buchdahl inspired metric is also proposed in the literature and tested with EHT shadow data as well \cite{Azreg-Ainou:2023qtf}. More recently, the free parameter $\tilde{k}$ of the theory has been tested using observed Einstein rings \cite{Maryam:2024wwd} and S2 star motion about the central black hole of the Milky Way \cite{Zhu:2024oxz2}. 

The first observations of gravitational waves by LIGO and the first image of a BH in the galaxy M87 have provided valuable opportunities to test GR in the strong-gravity regime and to explore various physical theories. Another promising method to test gravity near black holes is through quasi-periodic oscillations (QPOs), observed in the X-ray flux from black holes and neutron stars in X-ray binary systems, and detected as narrow peaks in the power density spectrum. The measurement of higher-frequency (HF) QPOs through the upper $\omega_{U}$ and lower $\omega_{L}$ frequencies with the relation $\omega_{U}:\omega_{L}=3:2$ \cite{Kluzniak01,Torok05,Remillard06ApJ} serves as a powerful tool for probing the nature of astrophysical black hole candidates in the strong gravitational field regime. It also helps astronomers understand the innermost regions of accretion disks and the masses, radii, and spin periods of white dwarfs, neutron stars, and black holes. It should be emphasized that HF QPOs are sourced by galactic microquasars and modeled to explain important aspects of the epicyclic motions on the accretion disks around black holes~\cite{Stuchlik13A&A,Stella99-qpo,Rezzolla_qpo_03a}. However, HF QPOs appear in a particular part of the accretion disk, still needing to seek its origin~\cite{Torok11}. It is worth noting that in recent years the QPOs within various resonance models have been extensively explored in different theories of gravity \cite{Titarchuk05qpo,Germana18qpo,Kolos15qpo,Stuchlik07qpo,Azreg-Ainou20qpo,Jusufi21qpo,Shaymatov20egb,Shaymatov22c,Ghasemi-Nodehi20qpo,Rayimbaev22qpo,Shaymatov23ApJ,Shaymatov23EPJP,Mustafa24Phys,2024PDU....4601561M,2024PDU....4601569D}. 

Astrophysical processes, such as X-ray observations (e.g. \cite{Bambi12a,Bambi16b}) and accretion disk dynamics (e.g., \cite{Abramowicz13}), combined with recent data (e.g., \cite{Fender04mnrs,Auchettl17ApJ,IceCube17b}) around compact objects, provide crucial tests for both general relativity (GR) and alternative theories of gravity. Analyzing the motion of the test particles around these objects offers another valuable approach to understanding their unique properties, and this motion can also be used to model accretion disks (e.g., \cite{Bambi17e,Chandrasekhar98}). These tests may reveal unexpected behaviors of compact astrophysical objects. Crucially, from an astrophysical perspective, the Buchdahl-inspired spacetime within $\mathcal{R}^2$ gravity is expected to significantly influence the time-like particle geodesics and accretion disk properties near a central object. From an astrophysical viewpoint, it is essential to examine the determining role of the Buchdahl-inspired spacetime in the aforementioned astrophysical processes around compact objects.
The \textit{special} Buchdahl-inspired metric is a solution to pure quadratic gravity. In regions near central objects, where the gravitational field is strong, Einstein's gravity may fail to describe adequately accretion processes. We believe that it is interesting to investigate the accretion matter and QPOs properties pertaining to this spacetime.

In this work, we intend to investigate the properties of the Buchdahl-inspired spacetime metric on the basis of epicyclic oscillations and the accretion disk in its surrounding environment. We first intend to analyze particle geodesics around Buchdahl-inspired spacetime, including conserved quantities for massive particles on ISCOs. We then derive general expressions for astrophysical epicyclic oscillations using the circular motion of massive particles. Astrophysical implications are explored by fitting higher-frequency QPOs from selected galactic microquasars to constrain spacetime parameters. We model accretion disks around Buchdahl-inspired spacetime, using ISCO parameters to define the inner edge. Finally, we intend to investigate the accretion disk's radiation properties and the redshifted image and intensity of a lensed accretion disk.

The plan of this paper is as follows. In Sec.~\ref{Sec:metric}, we briefly discuss the \textit{special} Buchdahl-inspired metric, which is followed by the formalism used to model the morion particles in this solution, describing the Buchdahl-inspired object.  In Sec.~\ref{Sec:QPO}, we derive the general expressions of frequencies for the QPOs and explore the Buchdahl parameter on the upper and lower frequencies by considering HF QPOs of two selected microquasars with their masses and frequencies.  We then consider the accretion disk with its radiative properties and explore redshifted image and intensity of a lensed accretion disk around the Buchdahl-inspired spacetime in Sec~\ref{Sec:accr_disk}. Finally, we provide a conclusion of our findings in Sec. \ref{Sec:con}. We use a spacetime metric signature $(–, +, +, +)$ and the system of units where $c=G=M=1$ throughout the paper.

\section{\label{Sec:metric} The \textit{special} Buchdahl-inspired metric }

The field equations of purely $\mathcal{R}^2$ gravity in vacuum are \cite{Nguyen:2022blj}
\begin{equation}
\mathcal{R}\,\Bigl(\mathcal{R}_{\mu\nu}-\frac{1}{4}g_{\mu\nu}\mathcal{R}\Bigr)+g_{\mu\nu}\square\,\mathcal{R}-\nabla_{\mu}\nabla_{\nu}\mathcal{R}=0,
\end{equation}
which contain fourth derivatives of the metric components $g_{\mu\nu}$ in $\square\,\mathcal{R}$ and $\nabla_{\mu}\nabla_{\nu}\mathcal{R}$. 

In our investigation, we will focus on the asymptotically flat vacuum solution outside of a static and spherically symmetric mass source. 

The \textit{special} Buchdahl-inspired metric, characterized by a static, spherically symmetric vacuum configuration and a vanishing cosmological constant, was introduced  in~\cite{Nguyen:2022blj}
\begin{eqnarray}\label{metric}
ds^2&=&\Big|1-\frac{r_s}{r}\Big|^{\tilde{k}}\Big\{-\left(1-\frac{r_s}{r}\right)dt^2+\left(\frac{\rho(r)}{r}\right)^4\frac{dr^2}{1-\frac{r_s}{r}}\nonumber\\&&+\left(\frac{\rho(r)}{r}\right)^2 r^2 d \Omega^2\Big\},
\end{eqnarray}
where the function $\rho(r)$ is given by virtue of
\begin{equation}\label{rho}
\left(\frac{\rho(r)}{r}\right)^2=\frac{\zeta^2\left|1-\frac{r_s}{r}\right|^{\zeta-1}}{\left(1-s\left|1-\frac{r_s}{r}\right|^{\zeta}\right)^2}\left(\frac{r_s}{r}\right)^2,
\end{equation}
$\tilde{k}$ and $\zeta$ are dimensionless parameters, which are defined by $\tilde{k}=k/r_s$ and $\zeta=\sqrt{1+3\tilde{k}^2}$, and $s=\pm 1$ denotes the signum of $1-\frac{r_s}{r}$. Here $\tilde{k}$ is a new (Buchdahl) parameter with higher-derivative characteristic, and $r_s$ plays the role of a Schwarzschild radius. If $\tilde{k}=0$, $\rho(r)=r$ and Eq.~\ref{metric} recovers the  Schwarzschild metric. It is worth noting that the \textit{special} Buchdahl-inspired metric lies at the intersection between the Buchdahl-inspired metric family and the null-Ricci-scalar metric family. $\mathcal{R}_{\mu\nu} \neq 0$ and $\mathcal{R}\neq 0$ for the Buchdahl-inspired metric, while $\mathcal{R}_{\mu\nu}\neq0$ and $\mathcal{R}=0$ for the \textit{special} Buchdahl-inspired metric. Since $\mathcal{R}\propto \Lambda$~\cite{Zhu:2024oxz}, where $\Lambda$ is the cosmological constant, we have $\Lambda \neq 0$ for the Buchdahl-inspired metric and $\Lambda = 0$ for the \textit{special} Buchdahl-inspired metric ($k \in \mathbb{R}$ for both metrics)~\cite{Nguyen:2022blj,Zhu:2024oxz,Nguyen24}. If $\Lambda$ and $k$ vanish, we obtain the Schwarzschild spacetime.
\vskip4pt 
Metric~\eqref{metric} is a solution to pure quadratic gravity. In regions near central objects, where the gravitational field is strong, Einstein's gravity may fail to describe adequately accretion processes. We believe that it is interesting to investigate the accretion matter and QPOs properties pertaining to the spacetime~\eqref{metric}. Other physical and geometric properties of the spacetime~\eqref{metric}, with astrophysical applications, have been discussed in a couple of papers to which we refer the interested reader for more details~\cite{Nguyen:2022bmj,Nguyen:2023kwr,Azreg-Ainou:2023qtf,Zhu:2024oxz,Zhu:2024oxz2,Nguyen:2023qux,Nguyen:2022bmj}.

\begin{figure}[!htb]
\centering
\includegraphics[scale=0.5]{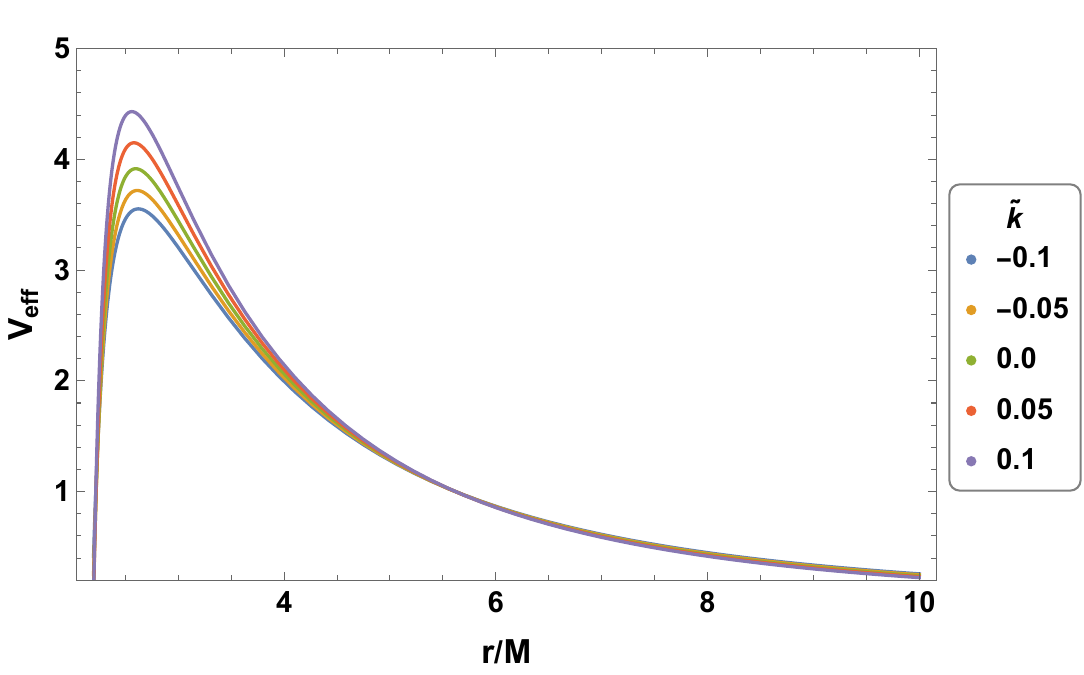}
\caption{Radial dependence of effective potential $V_{eff}$ of massive particle for the different values of parameter $\tilde{k}$.}
\label{fig:effektiv}
\end{figure}
\begin{figure*}[!htb]
\centering
\includegraphics[scale=0.45]{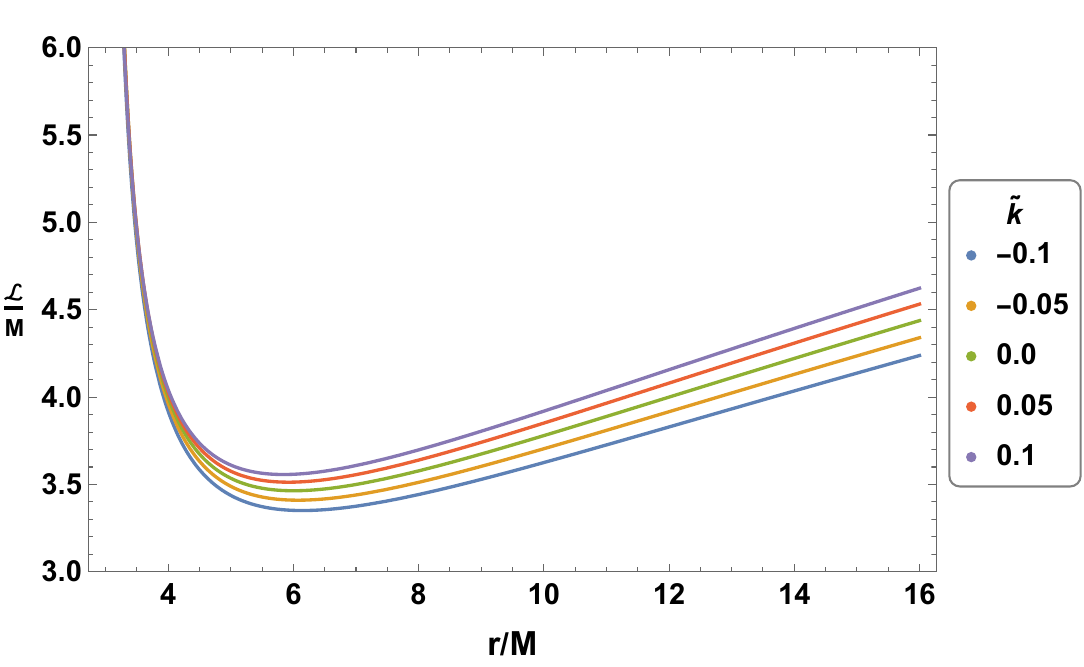}
\includegraphics[scale=0.45]{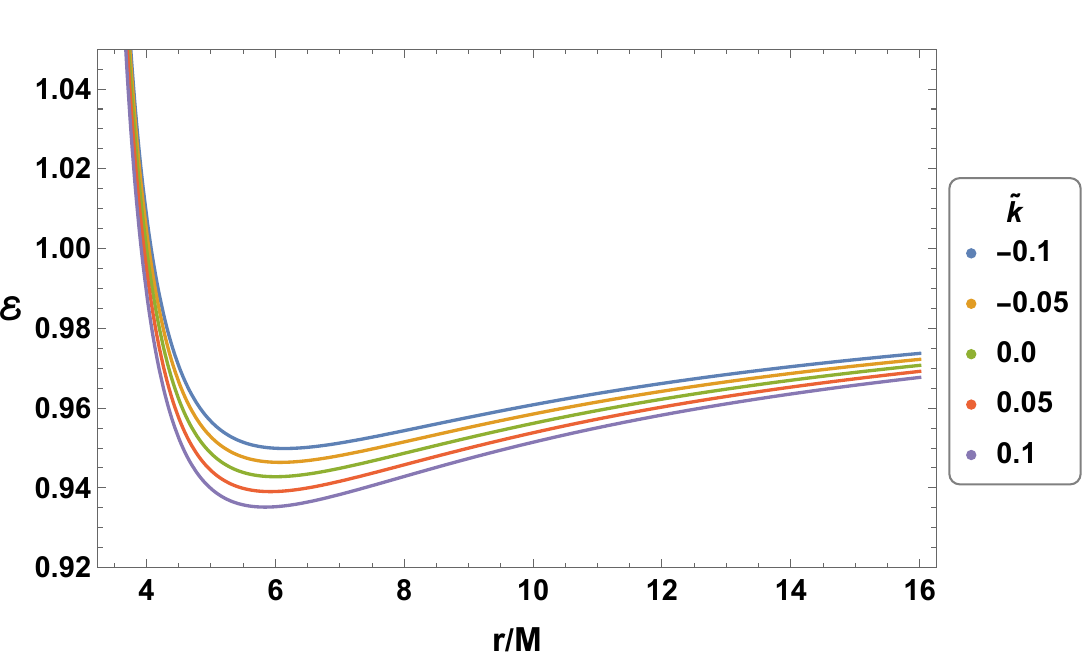}
\caption{Radial dependence of ${\cal{L}}$ and ${\cal E}$ for the different values of $\tilde{k}$ parameter.}
\label{fig:energy}
\end{figure*}
\begin{figure}[!htb]
\centering
\includegraphics[scale=0.5]{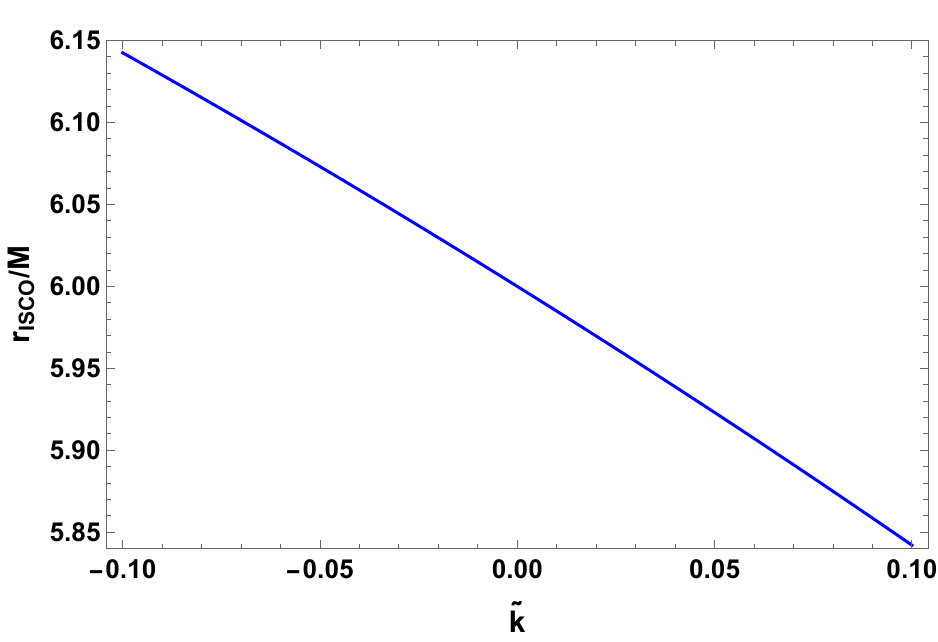}
\caption{Dependence of ISCO radius from $\tilde{k}$ parameter}
\label{fig:isco}
\end{figure}



Further, we investigate the test particle motion around the Buchdahl-inspired spacetime. To determine the trajectories of the test particles, we apply the principle of least action with a Lagrangian for a test particle with mass $m$ given by
\begin{align}
L=\frac{1}{2}g_{\alpha\beta}u^{\alpha}u^{\beta} \ , \qquad u^{\alpha}=\frac{dx^{\alpha}}{d \tau}
\end{align}
where $x^{\alpha}$ and $u^{\alpha}$ are the test particle's coordinates and four-velocity, respectively. $\tau$ is the proper time. Using the Lagrangian, we can write the conservative quantities of the motion as
\begin{eqnarray}\label{E and L}
-{\cal{E}}&=&\frac{\partial L}{\partial u^t}=g_{tt}\frac{dt}{d\tau}\ , \nonumber\\
{\cal{L}}&=&\frac{\partial L}{\partial u^{\phi}}= g_{\phi\phi}\frac{d\phi}{d\tau}\ .
\end{eqnarray}
where ${\cal{E}}$ and ${\cal{L}}$ are the energy and the angular momentum per unit mass respectively. One can find the $t$ and $\phi$ components of the four-velocity from Eq.~\ref{E and L} as
\begin{eqnarray}\label{fourvelocity}
u^{t}=-\frac{{\cal E}}{g_{tt}} ,\
u^{\phi}=\frac{{\cal L}}{g_{\phi\phi}}.
\end{eqnarray}
Using the normalization condition for test particle $g_{\alpha\beta}u^{\alpha}u^{\beta}=-1$, we can find the following equation 
\begin{equation}
g_{rr}\dot{r}^2+g_{\theta\theta}\dot{\theta}^2=V_\text{eff}(r,\theta),
\end{equation}
where $V_\text{eff}(r,\theta)$ is the effective potential. It can be written as~\cite{Misner73}
\begin{equation}\label{effektive}
V_\text{eff}(r)=-1+\dfrac{\mathcal{E}^2 g_{\phi \phi}+\mathcal{L}^2 g_{tt}}{-g_{tt}g_{\phi \phi}}\, ,
\end{equation}
\begin{figure*}[!htb]
\centering
\includegraphics[scale=0.37]{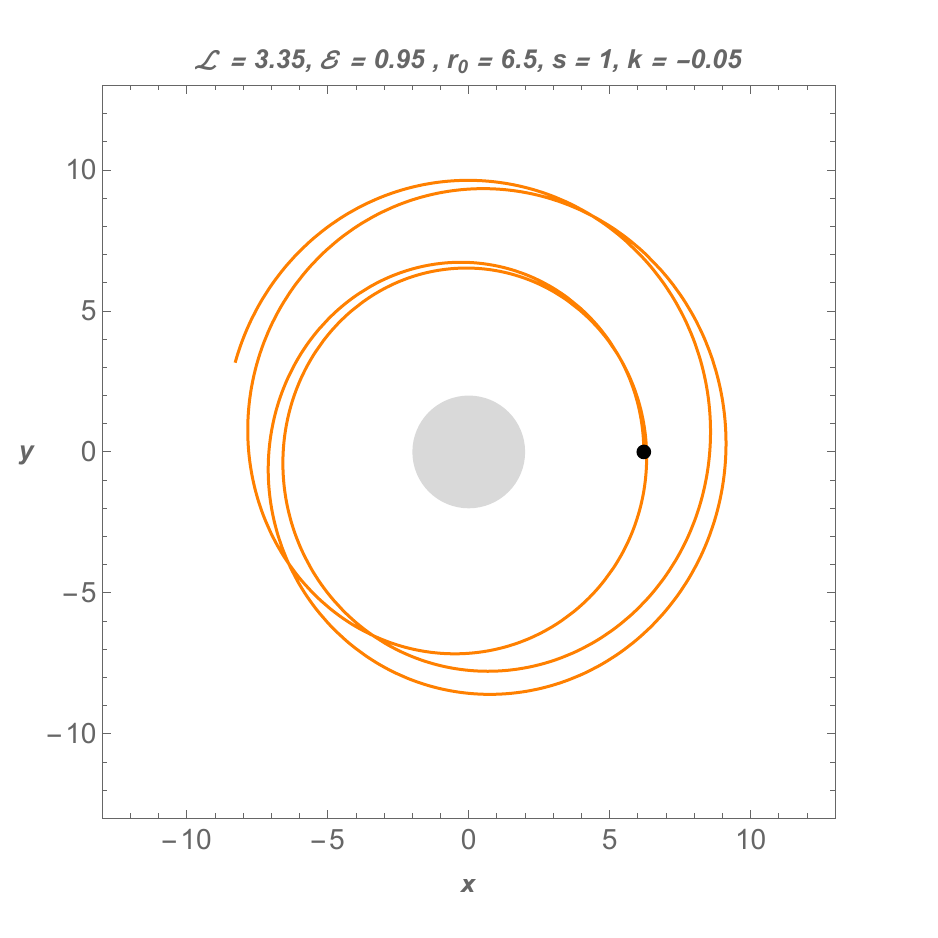}
\includegraphics[scale=0.37]{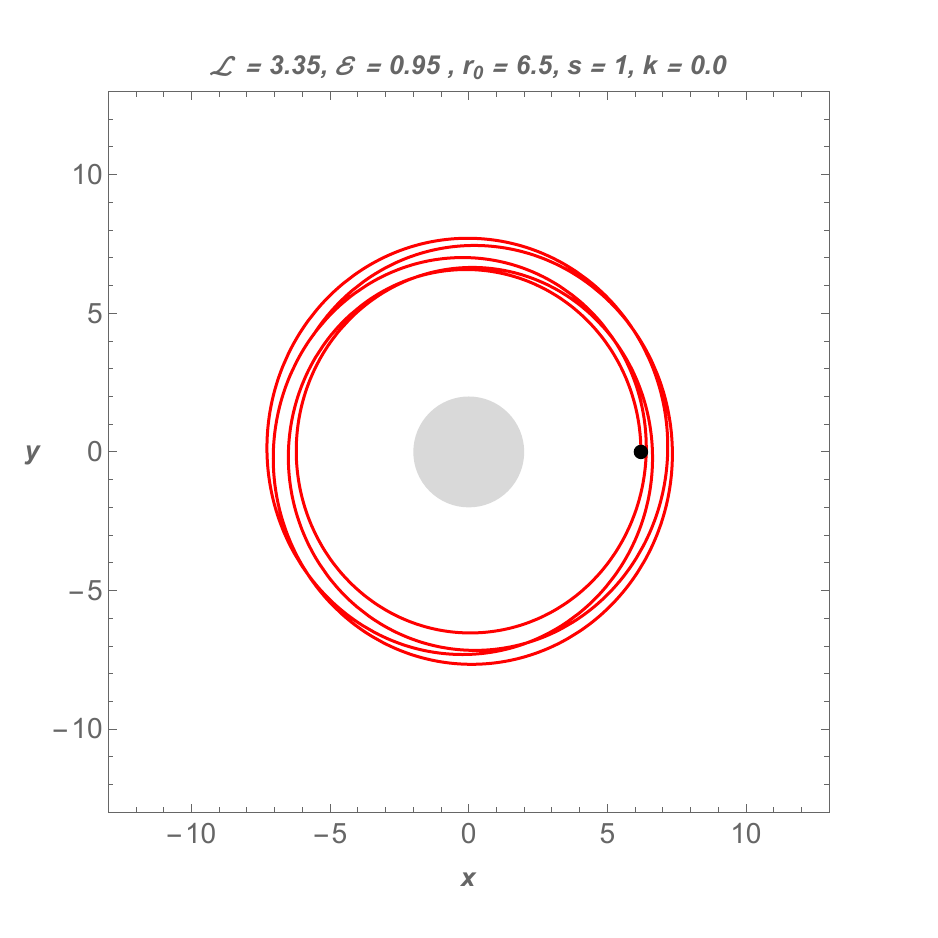}
\includegraphics[scale=0.37]{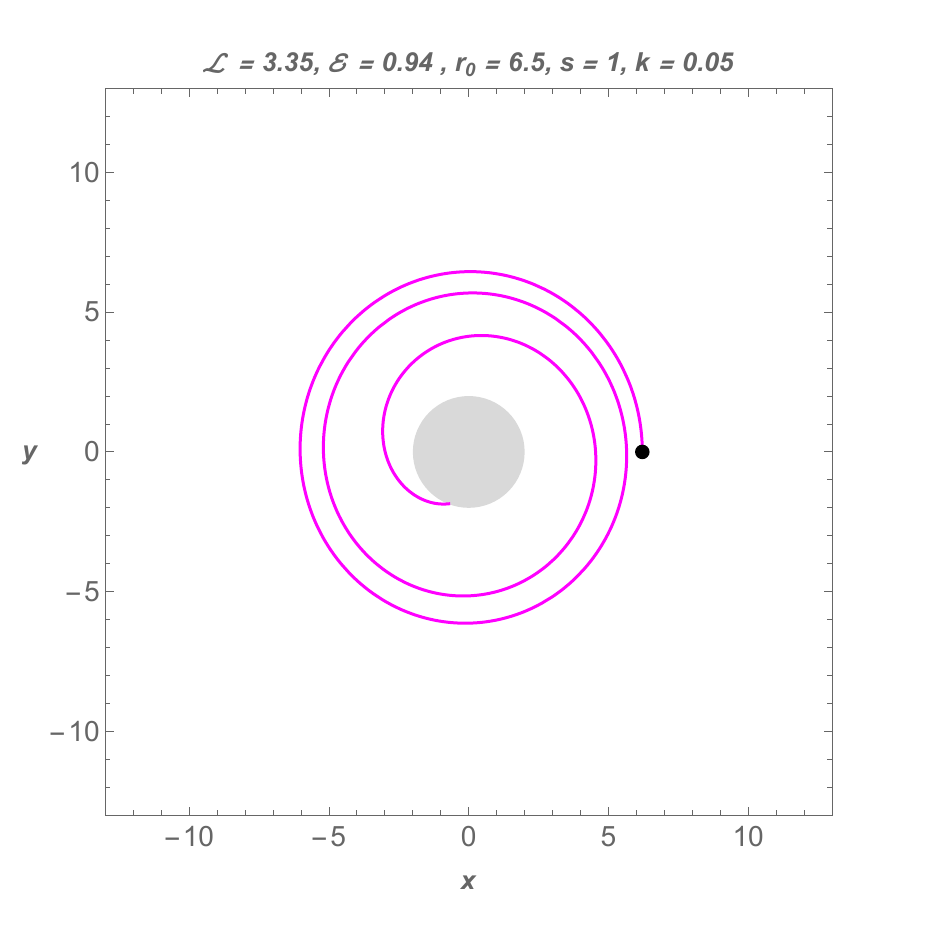}
\caption{The trajectories of the test particles in the vicinity of the Buchdahl-inspired spacetime for different values of the angular momentum and Buchdahl parameters. Here $\cal{L}$ and $\cal{E}$ are the specific angular momentum and the specific energy of the test particles, respectively.}
\label{fig:trajectory}
\end{figure*}
For circular motion, these two conserved quantities can be expressed using the metric tensors as~\cite{bambi2018introduction,universe10120451,sym14091765}
\begin{equation}\label{energy}
\mathcal{E}=-\dfrac{g_{tt}}{\sqrt{-g_{tt}-\Omega^2 g_{\phi \phi}}}\, , 
\end{equation}
and
\begin{equation}\label{angular}
\mathcal{L}=\dfrac{\Omega g_{\phi \phi}}{\sqrt{-g_{tt}-\Omega^2 g_{\phi \phi}}}\, . 
\end{equation}
Similarly, the orbital angular velocity of the test particle is also given by \cite{Shaymatov22c,Shapiro83,Shaymatov22a}
\begin{equation}
\Omega=\dfrac{d \phi}{d t}=\sqrt{-\dfrac{g_{tt,r}}{g_{\phi \phi,r}}}\ .
\end{equation}
Here, it is important to highlight that we are limiting the motion to the equatorial plane, denoted as $\theta=\pi/2$. With the Eqs.~\ref{metric}, \ref{rho}, and \ref{effektive}, we can plot the radial dependence of the effective potential for various values of the parameter $\tilde{k}$ as depicted in Fig.~\ref{fig:effektiv}.
Fig.~\ref{fig:energy} demonstrates the radial dependence of energy $\cal E$ and angular momentum $\cal L$ of test particles along circular stable orbits around a Buchdahl-inspired spacetime. We can also see that the values of the angular momentum $\cal L$ increase with increasing the parameter $\tilde{k}$. 
Moreover, we explore the radius of the innermost stable circular orbit (ISCO).  To consider ISCO radius one needs to apply the following general conditions:
\begin{equation}\label{effekt}
\begin{cases}
\partial_rV_\text{eff}=0,\\
\partial_{rr} V_\text{eff}=0.
\end{cases}
\end{equation}
We cannot get an analytical expression for the radius $r_{ISCO}$ from the above conditions. However, we present the radius of ISCO numerically using the plot for various values of the parameter $\tilde{k}$ in Fig.~\ref{fig:isco}. From this figure, one can obtain information about the dependence of the ISCO radius on the $\tilde{k}$. Particularly, ISCO radius decreases as $\tilde{k}$ increases. 
In addition, the trajectory of test particles near the Buchdahl-inspired spacetime was demonstrated in Fig.~\ref{fig:trajectory}. Here, the value of the Buchdahl parameter $\tilde{k}$ was increased from $-0.05$ to $0.05$. 
We can see from Eqs.~\ref{energy} and~\ref{angular} that ${\cal E}$ and ${\cal L}$ diverge when their denominator vanishes. This occurs at the radius of the photon sphere $r_{ph}$.
\begin{equation}\label{ph}
g_{tt}+\Omega^2 g_{\phi\phi}=0.
\end{equation}
One can find the radius of the photon sphere from Eq.~\ref{ph}. 
\begin{figure*}[!htb]
\centering 
\includegraphics[width=0.4\textwidth]{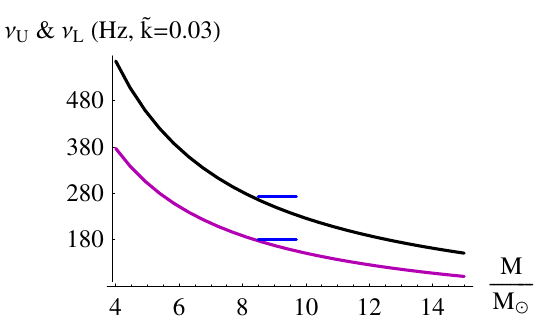}
\includegraphics[width=0.4\textwidth]{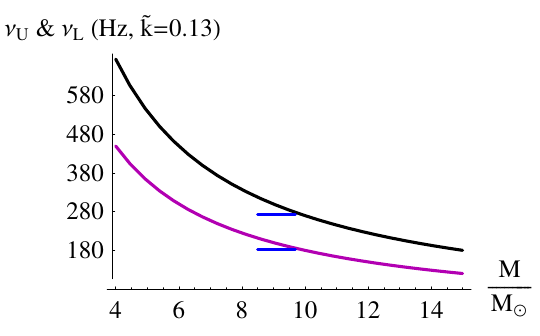}
\caption{Curve fit to the data of the XTE J1550-564 galactic microquasar; see Eq.~(\ref{XTE}). The black curves represent $\nu_U=\nu_\theta+\nu_r$, the purple curves represent $\nu_L=\nu_\theta$, and the blue lines represent the uncertainty on the mass of the XTE J1550-564 microquasar. \label{fig:xte}}
\end{figure*}
\begin{figure*}[!htb]
\centering 
\includegraphics[width=0.4\textwidth]{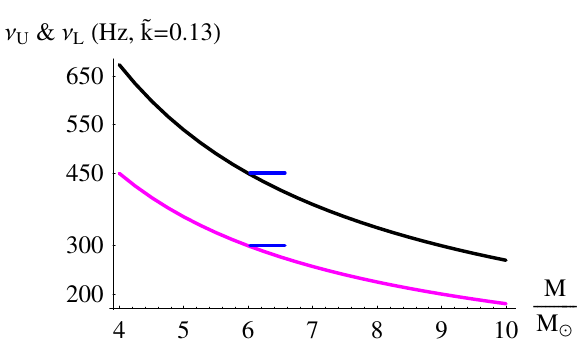}
\includegraphics[width=0.4\textwidth]{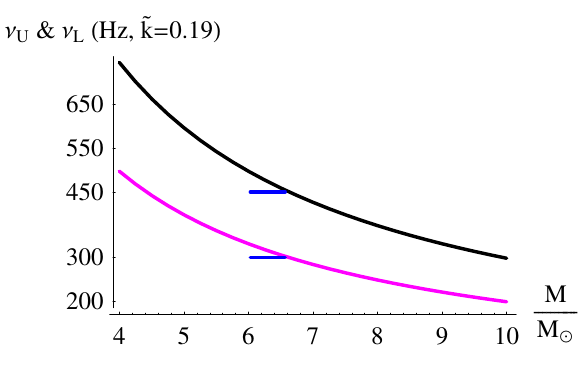}
\caption{Curve fit to the data of the GRO J1655-40 galactic microquasar with $\nu_U=450\pm 3$ Hz, $\nu_L=300\pm 5$ Hz and $M/M_\odot=6.30\pm 0.27$. The black curves represent $\nu_U=\nu_\theta+\nu_r$, the magenta curves represent $\nu_L=\nu_\theta$, and the blue lines represent the uncertainty on the mass of the GRO J1655-40 microquasar. \label{fig:gro}}
\end{figure*}

\section{Epicyclic oscillations around a special Buchdahl-inspired spacetime }\label{Sec:QPO}
In many applications, circular motion faithfully represents the trajectories of in-falling matter in accretion processes. As the in-falling matter reaches stable circular orbits, it undergoes quasi-periodic oscillations (QPOs) around the circular path in the plane containing this path, say the plane $\theta=\pi/2$, and perpendicular to it. These two components of epicyclic motion have observational frequencies~\cite{Torok05,McClintock11,Barret05,Belloni12,Kotrlova08,Strohmayer01ApJ,Shafee06ApJ,Mustapha20}.

For the general spherically symmetric metric,
\begin{equation}\label{q1}
ds^2=g_{tt}(r)dt^2+g_{rr}(r)dr^2+g_{\theta\theta}(r)d\Omega^2\,,
\end{equation}
the frequencies of the radial and vertical (polar) epicyclic motion have been derived in many references~\cite{Aliev81,Mustapha20,Mustapha19,Shaymatov22c} and they are given, respectively, by
\begin{align}
\label{q2}& \nu_r =\frac{1}{2\pi}\sqrt{\frac{2 g_{\theta\theta} (g_{tt}')^2-2 g_{tt} g_{tt}' g_{\theta\theta}'-g_{tt} g_{\theta\theta} g_{tt}''}{2 g_{tt} g_{rr} g_{\theta\theta}}+\frac{g_{tt}' g_{\theta\theta}''}{2 g_{rr} g_{\theta\theta}'}}\, ,\\
\label{q3} & \nu_\theta =\frac{1}{2\pi}\sqrt{\frac{-g_{tt}'}{g_{\theta\theta}'}}\, ,
\end{align}
where the prime notation denotes the derivative with respect to $r$. Here $\nu_r$ and $\nu_\theta$ are the frequencies of the perturbed circular motion as detected by an observer at spatial infinity, which are related to the local frequencies, $\Omega_r$ and $\Omega_\theta$, by $\nu_r=\Omega_r/(2\pi u^t)$ and $\nu_\theta=\Omega_\theta/(2\pi u^t)$ where $u^t$ and $u^\phi$ are the only non-vanishing components of the 4-velocity vector of the in-falling particle.

For the metric that is given in~\eqref{metric}, we introduce the function
\begin{equation}\label{f}
f(x)=1-\frac{r_s}{r}=1-\frac{2r_g}{(1+\tilde{k})r}=1-\frac{2}{(1+\tilde{k})x},
\end{equation}
where we have inserted the gravitational constant $G$ and the speed of light $c$ via $r_g=GM/c^2$, with $r=r_g x$ and $M$ is the mass of the black hole related to $r_s$ by $2r_g=(1+\tilde{k})r_s$~\cite{Azreg-Ainou:2023qtf}. The frequencies are given by
\begin{align}
\nu _r(x)&=\frac{c^3 (1+\tilde{k}) f^{1-\zeta }}{4 \pi  \sqrt{2} G M \zeta ^2 } \sqrt{\dfrac{(1-f^{\zeta })^3}{(1-\tilde{k}+\zeta )f^{\zeta }+\zeta +\tilde{k}-1}}\nonumber\\ &\times \left(4 (\tilde{k}^2-1) f^{\zeta }+(1+\tilde{k}) (2+\zeta ) (1-\tilde{k}+\zeta
) f^{2 \zeta }\right.\nonumber\\ &\left. +(1+\tilde{k}) (\zeta -2) (\tilde{k}+\zeta -1)\right)^{1/2}
\, , 
\\
\nu _{\theta }(x)&=\frac{c^3}{4 \pi  G M \zeta  } \frac{(1+\tilde{k})^{\frac{3}{2}} f^{1-\frac{\zeta }{2}}}{\sqrt{\dfrac{(1-\tilde{k}+\zeta )f^{\zeta }+\zeta +\tilde{k}-1}{(1-f^{\zeta
	})^3}}}\,.
\end{align}

The two peaks, upper $\nu_U$ and lower $\nu_L$, in the power spectra from the XTE J1550-564 galactic microquasar, its mass and the error bands are given by 
\begin{eqnarray}\label{XTE}
\frac{M}{M_\odot}=9.1\pm 0.6\, ,\;
\nu_U=276\pm 3 \text{ Hz},\;\nu_L=184\pm 5 \text{ Hz}\, ,
\end{eqnarray}
where $M_\odot$ is the mass of the sun. We have dropped the rotation parameter and its error band because our solution~\eqref{metric} is static. We aim to justify these two peaks using the static metric~\eqref{metric}, that is, we assume that the XTE J1550-564 galactic microquasar is modeled by the metric~\eqref{metric}. These two values of the QPOs, $\nu_U$ and $\nu_L$, are most certainly due to the phenomenon of resonance~\cite{Abramowicz03,Horak06,Rebusco04,Banerjee22,Deligianni21} which occurs in the vicinity of the ISCO, where the accreting particles perform radial and vertical oscillations around almost circular orbits. The most common models for resonances are parametric resonance and forced resonance. In parametric resonance we take $\nu_U = \nu_\theta,\, \nu_L =\nu_r$ with $\nu_\theta/\nu_r=n/2$ ($ n\in \mathbb{N}^+$). Since $\nu_\theta>\nu_r$ in the vicinity of ISCO, where accretion occurs and QPO resonance effects take place, the lower possible value of $n$ is 3. Forced resonance occurs in accretion processes when the central object (described here by the Buchdahl-inspired spacetime) becomes a source of possible perturbations for the accretion disk. These perturbations are due to couplings of the physical properties of the central object with the disk. Forced resonance has many variants~\cite{Banerjee22,Deligianni21,Shaymatov23ApJ} and in this work we use the model~\cite{SK2016}
\begin{eqnarray}\label{fr}
\nu_U=\nu_\theta+\nu_r, \qquad \nu_L=\nu_\theta\, .
\end{eqnarray}
Further discussions on the epicyclic resonance and its variants and relativistic precession models are available in the scientific literature~\cite{Banerjee22,Deligianni21,SK2016,testing24}.

We expect moderate accuracy in constraining the parameter $\tilde{k}$ by the QPOs due to two main reasons. First, we have neglected rotation of the microquasar due to the absence of an analytic rotating counterpart of the metric~\eqref{metric} that can faithfully represent the external geometry of XTE J1550-564 (all we have is a semi-analytic rotating metric~\cite{Azreg-Ainou:2023qtf}). Second, the data given in~\eqref{XTE} itself lack accuracy [including the rotation parameter not shown in~\eqref{XTE}]. Some constraints on the value of $\tilde{k}$ have been obtained in the literature~\cite{Azreg-Ainou:2023qtf,Tao23}, to mention the constraint $-0.155\leq\tilde{k}\leq 0.004$ obtained in~\cite{Azreg-Ainou:2023qtf} after modeling the central black hole M87{*} by the semi-analytic rotating solution counterpart of metric~\eqref{metric} and investigating its shadow. 

In Fig.~\ref{fig:xte} we plot the frequencies $\nu_U$ (black curves), $\nu_L$ (purple curves) versus $M/M_\odot$ constraining $x$ by $\nu_U/\nu_L=276/184=3/2$, and the uncertainty on the mass of the XTE J1550-564 microquasar (horizontal blue lines). In this graphical method, we aim to constrain $\tilde{k}$ by looking for plots of ($\nu_U,\,\nu_L$) that intersect the plots representing the uncertainty on the mass of the microquasar (mass-error band). For $\tilde{k}<0.03$ only one of the two plots ($\nu_U,\,\nu_L$) or none intersects the mass-error band. The intersection occurs for $0.03\leq\tilde{k}\leq 0.13$, as shown in both panels of Fig.~\ref{fig:xte}.

There are two peaks in the power spectra from the GRO J1655-40 galactic microquasar, which occur at $\nu_U=450\pm 3$ Hz and $\nu_L=300\pm 5$ Hz with $M/M_\odot=6.30\pm 0.27$. For this microquasar, the results are less satisfactory as the intersections of the two plots $\nu_U$ (black), $\nu_L$ (magenta) with the mass-error-band (blue) occur for $0.13\leq\tilde{k}\leq 0.19$, limits that are beyond the upper limit obtained by the shadow investigation~\cite{Azreg-Ainou:2023qtf}, as depicted in Fig.~\ref{fig:gro}.

The QPOs analysis allows us the conclude that the $\tilde{k}$ parameter should be taken to satisfy the inequalities $0.03<\tilde{k}<0.19$.

\section{The accretion disk with its radiative properties around the Buchdahl-inspired spacetime  
} \label{Sec:accr_disk}

In this section, we consider the accretion disk when the central object is described by the Buchdahl-inspired spacetime. Our main aim in this section is to see how the parameter $\tilde{k}$ affects the flux of the electromagnetic radiation, the temperature of the accretion disk, and the luminosity profile of the latter as compared with the Schwarzschild accretion case. We will let $\tilde{k}$ vary within the limits determined by the QPOs and other analyses~\cite{Azreg-Ainou:2023qtf,Zhu:2024oxz,Zhu:2024oxz2}. To this end, the accretion disk is assumed to be optically thick and geometrically thin around the Buchdahl-inspired spacetime \cite{Novikov:1973} in such a way that $h \ll R$, with $R$ being the radius of the disk. It is worth noting that there are other types of accretion disk models, such as a Bondi model~\cite{1952MNRAS.112..195B}. The Bondi model is radiatively inefficient because the energy is advected or lost via outflows rather than radiated. It also produces a yield faint, non-thermal spectra. Our study focuses on the Novikov-Thorne model of the accretion disk around the Buchdahl-inspired spacetime due to it produces a thermal spectrum and robustly fits thermal components in X-ray binaries~\cite{Davis_2006}. According to this thin-accretion disk model, the vertical entropy and pressure gradients of the disk's hydrodynamic equilibrium can be considered negligible in the accreting matter. This leads to the fact that the generated heat can not be accumulated in the accretion disk as it is efficiently cooled by the thermal radiation from the disk surface,  resulting in a stabilized thin disk with the inner edge located at the ISCO around the central object. The accretion disk's bolometric luminosity is then defined by~\cite{Bokhari20,Rayimbaev-Shaymatov21a}
\begin{eqnarray}
\mathcal{L}_{bol}=\eta \dot{M}c^2\, ,
\end{eqnarray}
$\dot{M}$ and $\eta$ being the rate of accretion matter and the energy efficiency of the accretion disk. Irrespective of astrophysical issues pertaining to the bolometric luminosity observation, it is increasingly important to measure it using theoretical analysis and various models. By defining energy efficiency, one can extract the highest energy from the accretion and can provide a deeper understanding of the remarkable nature of the accretion process, by which the rest-mass-accreting matter is transformed into electromagnetic radiation that can be emitted as the radiation rate of the photon energy from the disk surface~\cite{Novikov:1973,1974ApJ...191..499P}. Hence, the energy efficiency of the accretion disk can only be measured by the emitted photon energy at the ISCO on the disk \cite{sym14091765} and is given by $\eta=1-\mathcal{E}_{ISCO}$ (see, e.g. \cite{Bardeen73}). 

The expression above for the measured energy at the ISCO allows one to find the radiative efficiency $\eta$ for the emitted photon from the disk. Therefore, it is essential to determine the ISCO parameters, that is, $r_{ISCO}$, $\mathcal{E}_{ISCO}$, and $\mathcal{L}_{ISCO}$.

\begin{table}[ht!]
\centering
\begin{tabular}{|l|c|c|c|c|c|r|}
\hline
$\tilde{k}$   & ${r}_{ISCO}$ & $\mathcal{L}_{ISCO}$ & $\mathcal{E}_{ISCO}$ & $\eta$\%  \\
\hline
-0.10    & 6.1425   & 3.3509 & 0.94988 & 5.01171 \\
-0.05    & 6.0731   & 3.4100 & 0.94640 & 5.35924  \\
-0.02    & 6.0297   & 3.4431 & 0.94426 & 5.57359 \\
-0.01    & 6.0149   & 3.4537 & 0.94353 & 5.64608 \\
~0.00    & 6.0000   & 3.4641 & 0.94289 & 5.71910 \\
~0.01    & 5.9849   & 3.4743 & 0.94207 & 5.79266 \\
~0.02    & 5.9696   & 3.4842 & 0.94133 & 5.86677 \\
~0.05    & 5.9230   & 3.5130 & 0.93907 & 6.09255  \\
~0.10    & 5.8421   & 3.5570 & 0.93518 & 6.48106 \\
\hline
\end{tabular}
\caption{{Numerical values of the ISCO radius $r_{ISCO}$, specific angular momentum $\mathcal{L}_{ISCO}$ and specific energy $\mathcal{E}_{ISCO}$ in ISCO for test particles, the radiative efficiency of the accretion disk.    \label{Table1}}} 
\end{table}
\begin{figure}[!htb]
\includegraphics[width=0.45\textwidth]{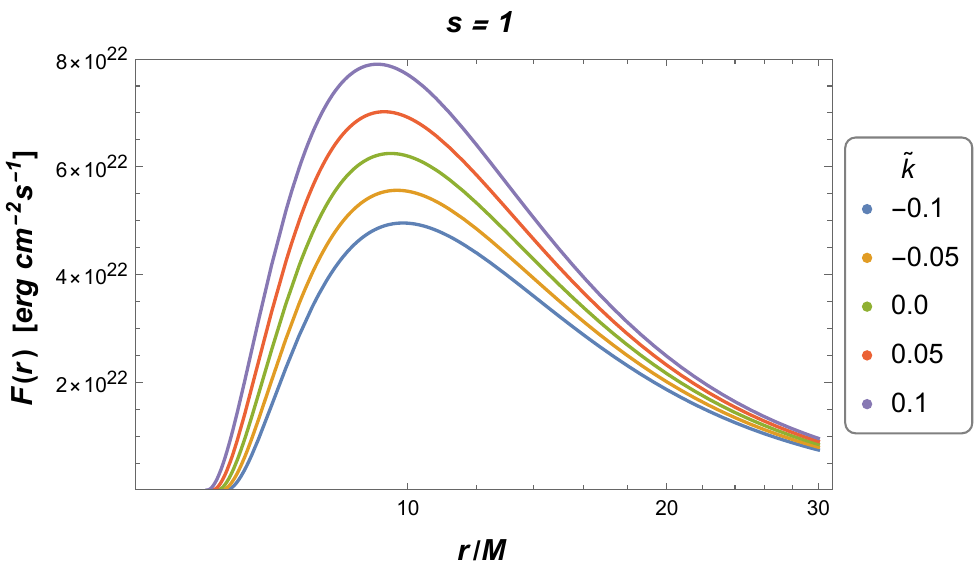}  
\caption{\label{fig:flux} Radial dependence of the flux of electromagnetic radiation of the accretion disk on different values of $\tilde{k}$ plotted for the Buchdahl-inspired spacetime.}
\end{figure}  
\begin{figure}[!htb]
\includegraphics[width=0.45\textwidth]{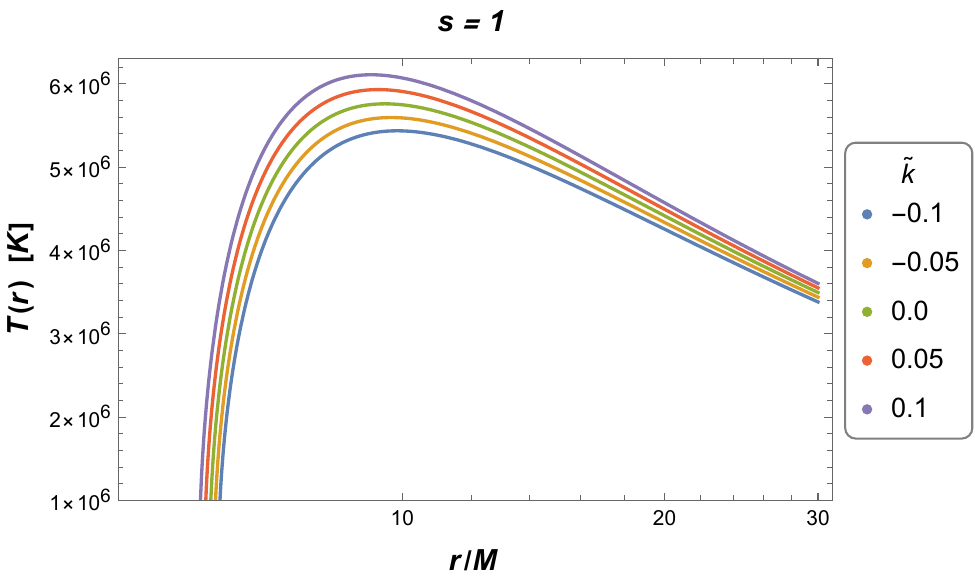}
\caption{\label{temperature1} Radial dependence of the temperature of the accretion disk for different values of $\tilde{k}$.}
\end{figure} 
\begin{figure}[!htb]
\centering 
\includegraphics[scale=0.5]{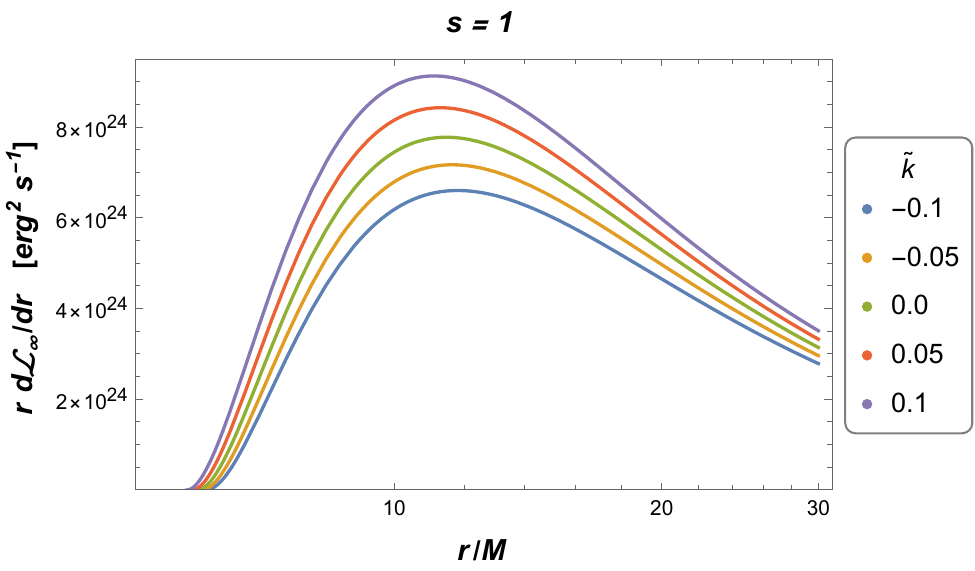}
\caption{The differential luminosity profile of the accretion disk for different values of the $\tilde{k}$ parameter.} 
\label{fig:luminosity}
\end{figure}
\begin{figure*}[!htb]
\centering 
\includegraphics[width=0.4\textwidth]{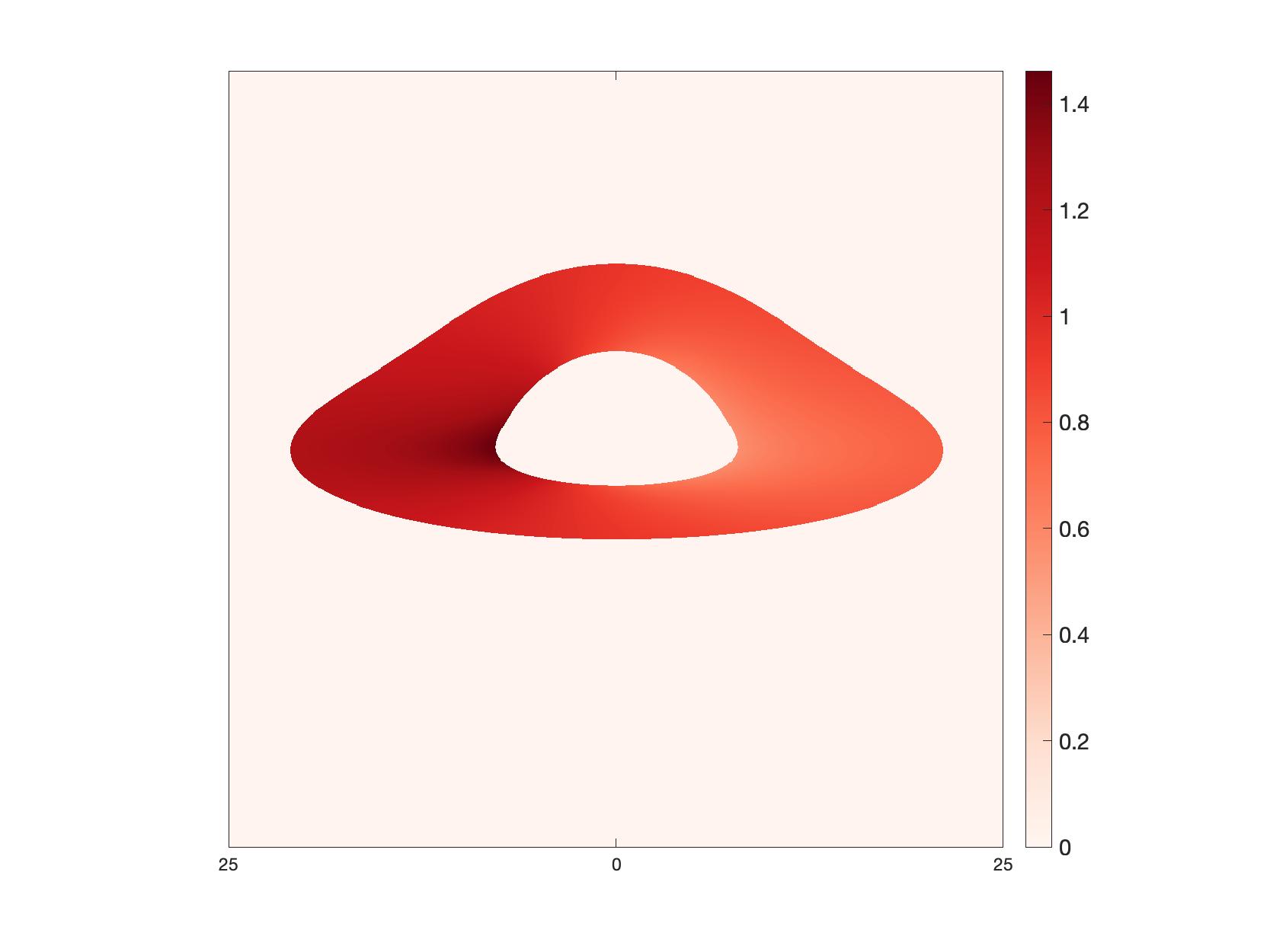}
\includegraphics[width=0.4\textwidth]{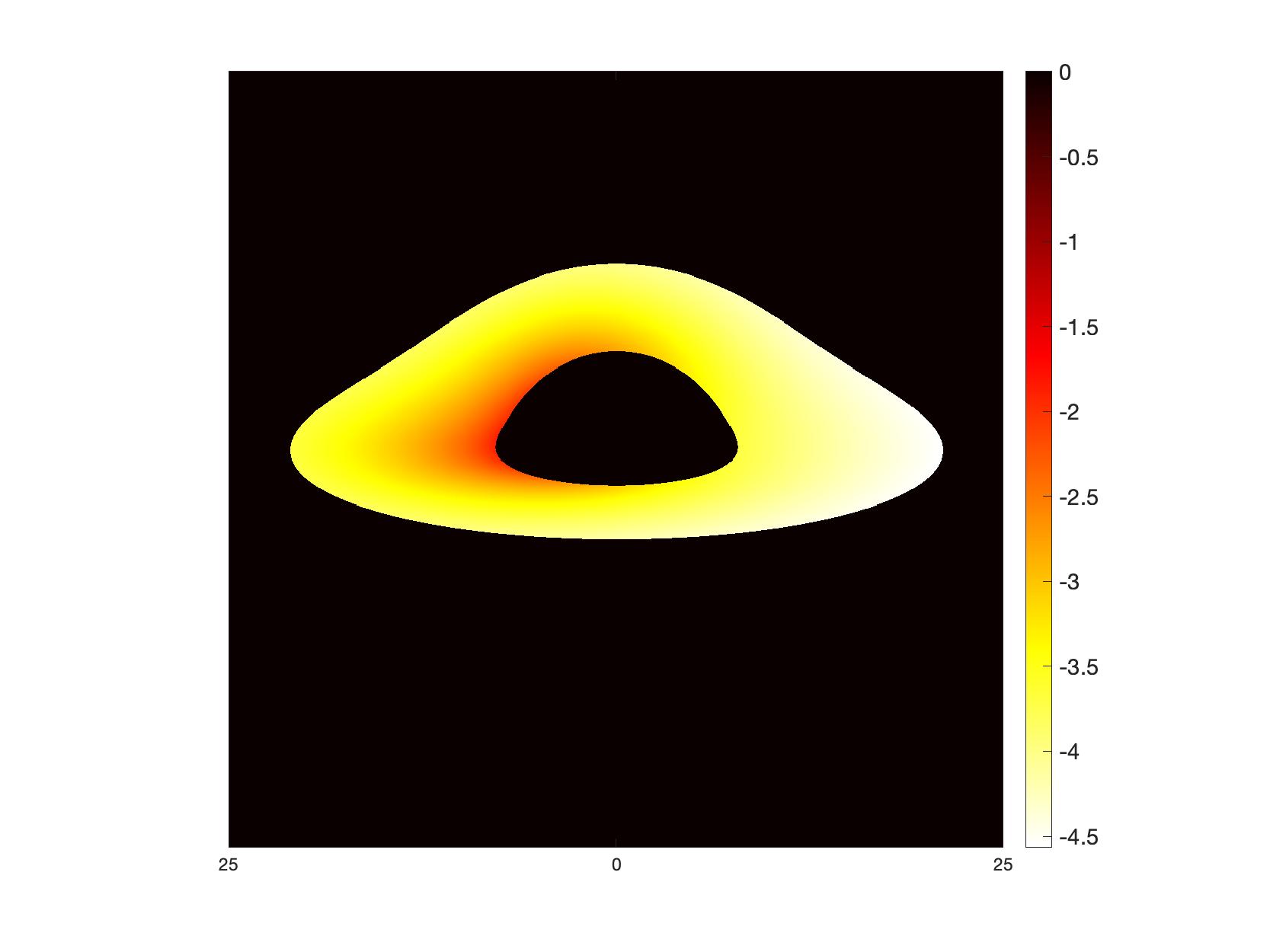}
\includegraphics[width=0.4\textwidth]{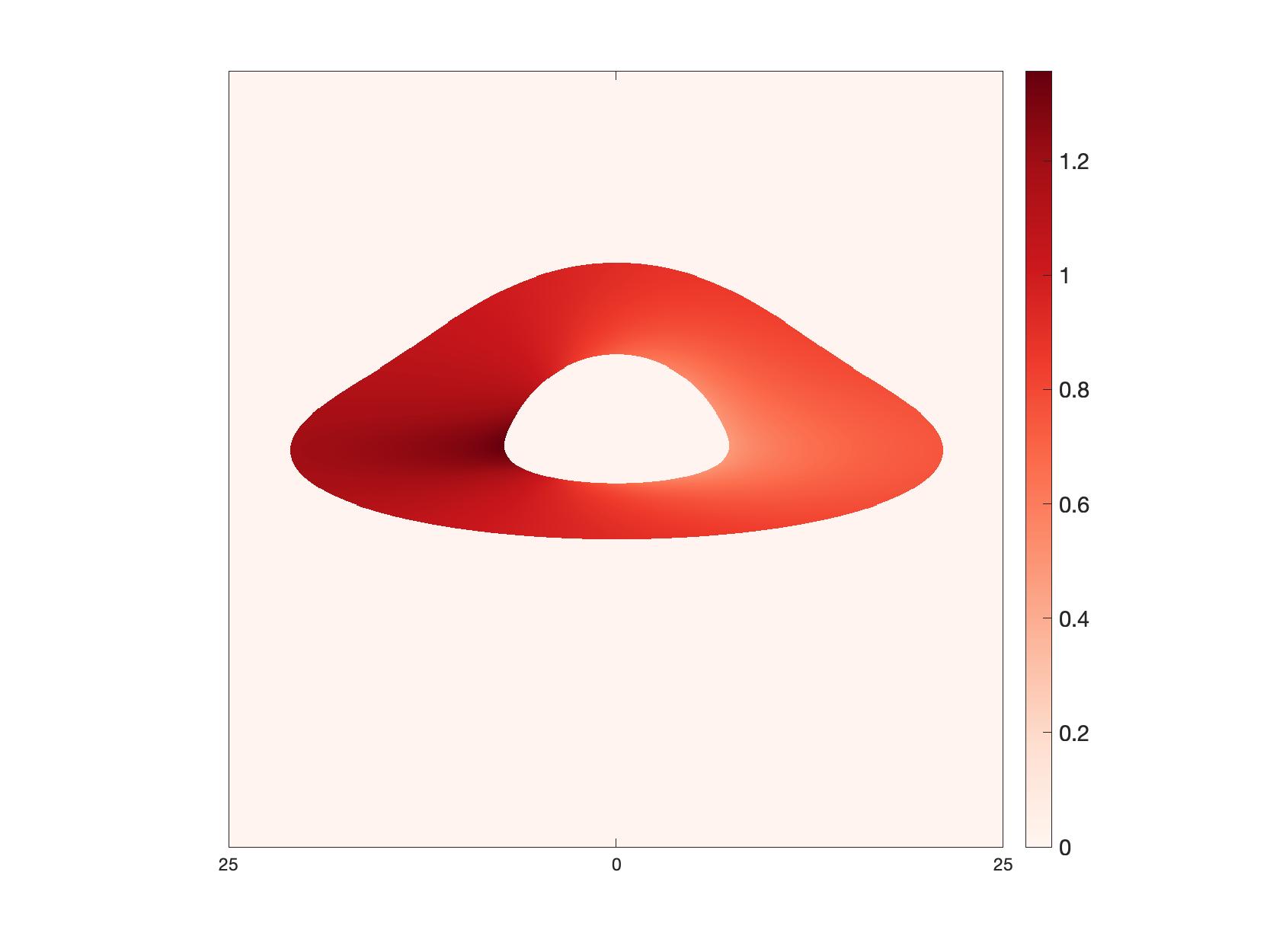}
\includegraphics[width=0.4\textwidth]{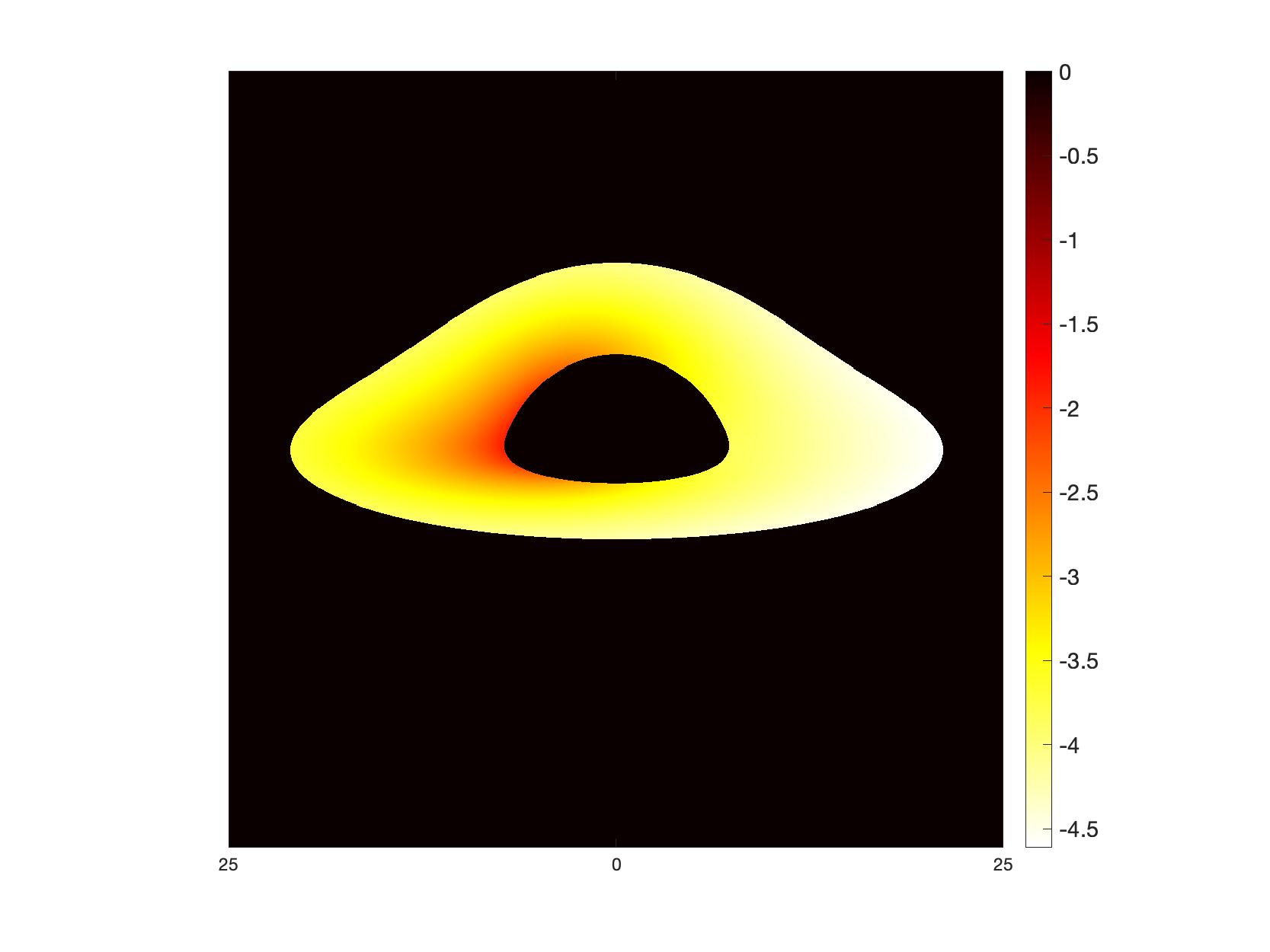}
\includegraphics[width=0.4\textwidth]{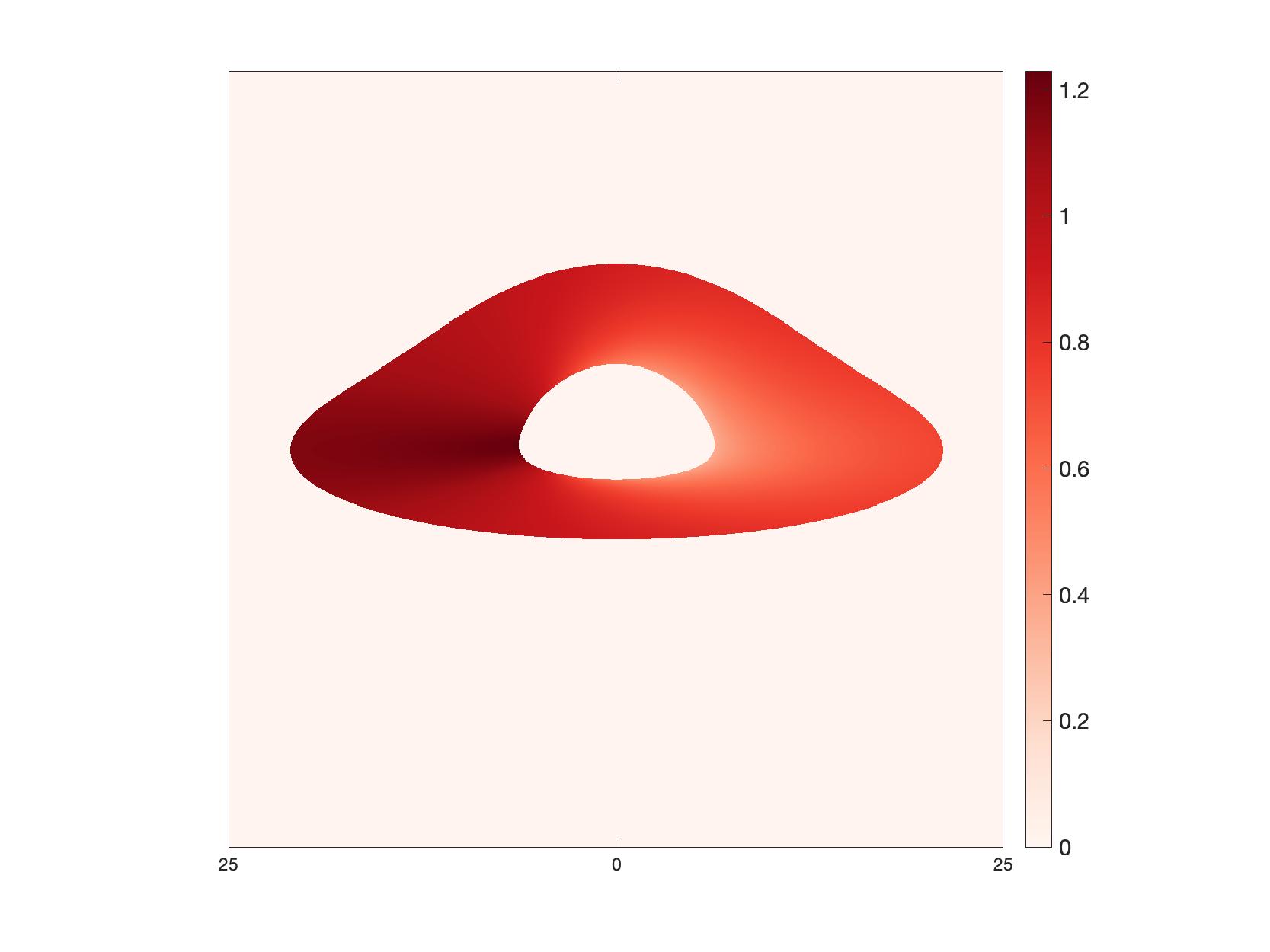}
\includegraphics[width=0.4\textwidth]{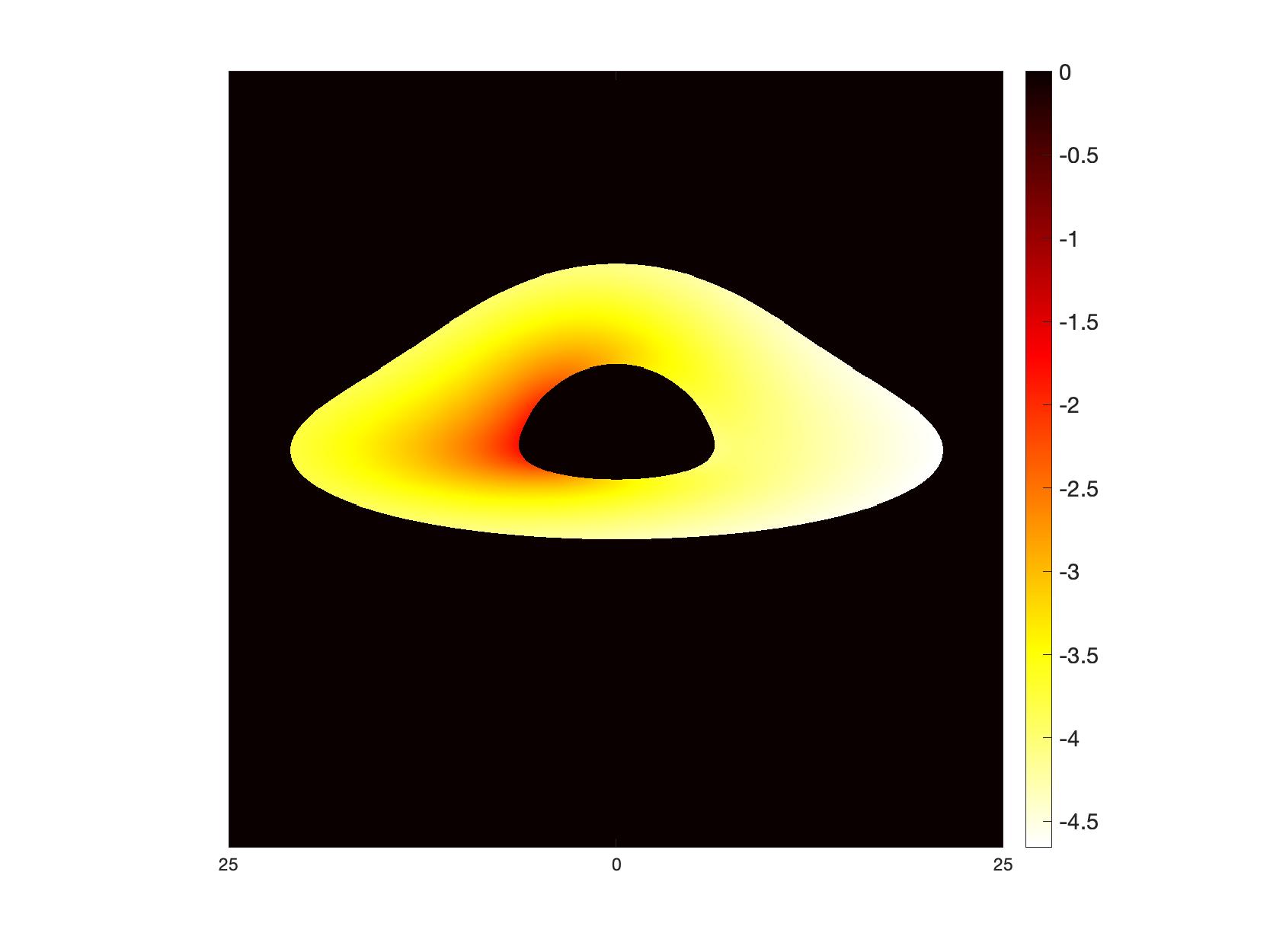}
\caption{Redshifted image (left panel) and intensity (right panel) of a lensed accretion disk around Buchdahl-inspired spacetimes with $\tilde{k}=-0.5,0,0.5$ from top to bottom. The inclination angle is set to $75^\circ$. The inner edge of disk is at $r_{ISCO}$ and the outer edge of disk is at $r=20M$} \label{fig:image}
\end{figure*}
Further, we study the flux produced by an accretion disk of the Buchdahl-inspired spacetime. We note that an accretion disk consists of gas and dust moving on stable orbits of a black hole or neutron star. The gas and dust can lose their energy and angular momentum because of the background geometry when they start moving around the compact object. As a result, the orbits of gas and dust move to the inner edge of the accretion disk so that they fall towards the compact object. This process ends up with accretion disk radiation, in which the gas and dust can heat up and emit radiation. We further expect that the accretion disk radiation can be affected by the Buchdahl-inspired spacetime parameters. The Buchdahl parameter $\tilde{k}$ is expected to influence the ionization process of the accretion disk. As a consequence of the ionization process, high-energy radiation can be emitted in the form of X-rays that can be detected. The radiation of the accretion disk around the Buchdahl-inspired spacetime is expected to be a potent test to explain valuable and remarkable aspects of the accretion disk and can also lead to serious implications from the observational point of view. The flux of the electromagnetic radiation can be defined by the following equation~\cite {Novikov:1973,Shakura:1972te,Thorne:1974ve,boshkayev2021accretiondiskluminosityblack,Boshkayev_2022,boshkayev2023luminosityaccretiondisksrotating,Jumaniyozov2025,Alloqulov2024}:
\begin{equation}\label{flux}
\mathcal{F}(r)=-\dfrac{\dot{M_0}}{4 \pi \sqrt{g}}\dfrac{\Omega_{,r}}{(E-\Omega L)^2} \int _{r_{ISCO}}^r (E-\Omega L) L_{,r} d r\, , 
\end{equation}
where $g$ is the determinant of the three-dimensional subspace $g=-g_{tt}g_{rr}g_{\phi \phi}$. We can define the quantity $\mathcal{F}(r)$ as a function of $\dot{M_0}$ that can be regarded as unknown and corresponds to the accretion rate of the disk mass. However, we can set $\dot{M_0}=1$ for further analysis. Using Eqs.~(\ref{energy}) and (\ref{angular}) we further derive the flux of electromagnetic radiation. However, it does turn out that it is complicated to derive the analytical expression of the flux, and thus we resort to it numerically. As can be observed in Fig.~\ref{fig:flux}, the maximal value of the flux increases under the effect of the parameter $\tilde{k}$. Moreover, we can write the flux of the black body radiation as $\mathcal{F}(r)=\sigma T^4$ with $\sigma$ as the Stefan-Boltzmann constant. The radial dependence of the disk temperature is shown in Fig.~\ref{temperature1} for various values of the parameter $\tilde{k}$. The radial dependence of the energy, defined by equation $E=kT$, is shown in the right panel of Fig.~\ref{temperature1} for different values of the parameter $\tilde{k}$. Note that $k$ refers to the Boltzmann constant. We can see from this figure that the values of both aforementioned quantities increase with increasing the parameter $\tilde{k}$. In addition, we investigate another important quantity. referred to as the differential luminosity. It can be expressed as follows:~\cite{Novikov:1973,Shakura:1972te,Thorne:1974ve,kurmanov2023accretiondiskhartlethornespacetime,kurmanov2024accretiondiskspropertiesregular,kurmanov2025acc}
\begin{equation}\label{Eq:luminosity}
\dfrac{d \mathcal{L}_{\infty}}{d \ln{r}}=4 \pi r \sqrt{g} E \mathcal{F}(r)\, .
\end{equation}
We assume that the radiation emission can be described by black body radiation. Keeping this in mind, one can then define the spectral luminosity $\mathcal{L}_{\nu,\infty}$ as a function of the radiation frequency $\nu$ at infinity as follows:~\cite{Boshkayev:2020kle,sym14091765,Shaymatov2023,Alloqulov_2024,Turimov:2024tvt,D_Agostino_2023}
\begin{equation}\label{luminosity2}
\nu \mathcal{L}_{\nu,\infty}=\dfrac{60}{\pi^3} \int_{r_{ISCO}}^{\infty} \dfrac{\sqrt{g} E}{M^2}\dfrac{(u^t y)^4}{\exp\Big[{\dfrac{u^t y}{(M^2 \mathcal{F})^{1/4}}}\Big]-1} dr\, ,
\end{equation}
with $y=h \nu /k T_{\star}$, where $h$ and $k$ refer to the Planck and Boltzmann constants, respectively.
In addition, $T_{\star}$ corresponds to the characteristic temperature and has a relation with the Stefan-Boltzmann law, that is, $\sigma T_{\star}= \dfrac{\dot{M}_0}{4 \pi M^2}$. The radial dependence of the differential luminosity is represented in Fig.~\ref{fig:luminosity}.  As can be observed, the value of differential luminosity is an increasing function of $\tilde{k}$.

To construct the images for a thin accretion disk around the Buchdahl-inspired spacetime, we utilize the ray-tracing techniques~\cite{1973ApJ...183..237C,1975ApJ...202..788C,10.1093/pasj/49.2.159,Chen_2015}. For the ray-tracing geometry, we define the set of Cartesian coordinates $(X_i,Y_j)$ representing the pixel on the image plane, located at a distance $D$ from the Buchdahl-inspired spacetime with inclination angle $i$. We then convert the coordinates $(X, Y,0)$ of a photon reaching the image plane into the coordinates $(r,\theta,\phi)$ in the spherical polar system used for the metric, employing the following relations:
\begin{eqnarray}
r&=&\sqrt{D^2+X^2+Y^2},\\
\cos \theta &=& \frac{D \cos i + Y \sin i}{r}, \\
\tan \phi &=& \frac{X}{D\sin\theta -Y\cos\theta}.
\end{eqnarray}
With these relations, we can integrate the geodesic equations backward in time from any detection point $(X_i, Y_j, 0)$ in the image plane of the distant observer to the emission point in the disk:
\begin{eqnarray}
\frac{d^2 x^\mu}{d\lambda^2}+\Gamma^\mu_{\nu\rho}\frac{dx^\nu}{d\lambda}\frac{dy^\rho}{d\lambda} = 0,
\end{eqnarray}
where $\lambda$ is an affine parameter. In the numerical algorithm, we integrate geodesic equations using the Runge-Kutta algorithm implemented in the ODE45 function of Matlab. The integration continues until the photon either reaches the accretion disk or approaches the event horizon too closely. Once on the accretion disk, we assume that the radiation emission follows a power law, given by:
\begin{equation}
I(\nu,\mu,r)\propto\frac{1}{r^n}\frac{\omega(\mu)}{\nu^{\Gamma-1}},
\end{equation}
where $\nu$ is the frequency of photon in the rest frame and $\mu$ is cosine of the angle between the photon’s 4 momentum and the upward disk normal measured by the comoving observer, $\omega(\mu)$ is the angular-dependence of the intensity profile taking from Chadrasekhar's book \cite{chandra:1960}, $\Gamma$ is a photon index and $n$ represent the radial steepness of the intensity profile. With the above source profile, The observed flux is
\begin{equation}
F_{\nu_{o}}=\int g^3 I^{source}_{\nu_e}d\Omega_o,
\end{equation}
where $\nu_e$ is the source frequency and $g\equiv\frac{\nu_o}{\nu_e}$ is the redshift factor.

In Fig.~\ref{fig:image} we plot the ray-traced redshifted image (left panel) and the intensity (right panel) of a lensed accretion disk at inclination angle $75^\circ$ for different values $\tilde{k}=-0.5,0,0.5$. The left panel of the figure shows that the Doppler blueshift in the left half-part of the plane can exceed the overall gravitational redshift. We also note that the Buchdahl-inspired spacetime significantly changes the shape and intensity profile of the disk. From the right panel of Fig.~\ref{fig:image}, the intensity is strongly concentrated in a small region (in red) near the Buchdahl-inspired spacetime and to the left of the accretion disk (where the source is approaching the observer). The hat-like structure shown in the figures is formed by bending the accretion disk's flat structure because of the Buchdahl-inspired spacetime's gravitational lensing effect. We found that the ray-traced images closely resemble the Schwarzschild case. However, the parameter $\tilde{k}$ significantly affects the size of the hollow region of the accretion disk, which is consistent with the results for \( r_{ISCO} \) at varying values of $\tilde{k}$.

\section{Conclusions}\label{Sec:con}

In this paper, we explored the Buchdahl-inspired spacetime and studied its effect on the test particle geodesics, astrophysical quasiperiodic oscillations and the accretion disk due to the accreting matter. It should be noted that the Buchdahl-inspired spacetime model supports various geometries and was also considered a modification of the Schwarzschild BH. Therefore, exploring distinguishing aspects of the Buchdahl-inspired spacetime would lead to the enhancement of our qualitative understanding of the nature of this spacetime and its crucial determining role and impact on the surrounding geometry and accretion disk. 

We began analyzing the time-like particle geodesics around the Buchdahl-inspired spacetime. To this end, we explored the radial dependence of the effective potential for various possible cases of the Buchdahl parameter $\tilde{k}$. The parameter $\tilde{k}$ was shown to enhance the potential barrier for massive particles around the Buchdahl-inspired spacetime. We showed that the ISCO radius diminishes as a consequence of an increase in the parameter $\tilde{k}$. Furthermore, we explored ISCO parameters such as the angular momentum $L_{ISCO}$ and the energy $\mathcal{E}_{ISCO}$ as a consequence of the influence of the Buchdahl parameter $\tilde{k}$. We showed that the ISCO angular momentum for massive particles increases, while the ISCO energy is reduced under the impact of the parameter $\tilde{k}$. This can be interpreted by the fact that a particle needs more angular momentum to remain stable as it was initially. 

Further, we considered the HF QPOs of two microquasars assuming that their masses and frequencies are in the middle of their band errors. This results in weak constraints on the Buchdahl parameter $\tilde{k}$. In contrast, we could consider the constraints on $\tilde{k}$ obtained by other investigations~\cite{Azreg-Ainou:2023qtf,Tao23} and use them to constrain the masses and/or the frequencies of the peaks and their bands, as we did in~\cite{Shaymatov23ApJ}. However, rotation remains an important ingredient that we lacked in our investigation due to the nonavailability of an analytical Buchdahl-inspired rotating metric.

Furthermore, we delved into the accretion disk of the Buchdahl-inspired spacetime and its radiation properties, such as flux, temperature, and differential luminosity. We provided corresponding figures of their profiles to support our analysis. From the results, we observed that the radiation properties, including the electromagnetic radiation flux, temperature, and differential luminosity of the accretion disk around the Buchdahl-inspired spacetime, are enhanced as a consequence of the influence of the Buchdahl parameter $\tilde{k}$. Additionally, we studied the redshifted image and intensity of a lensed accretion disk around the Buchdahl-inspired spacetime and plotted the hat-like structure formed by bending the accretion disk's flat structure because of gravitational lensing effects around the central object. We demonstrated the profile of the ray-traced redshifted image and the intensity of the accretion disk as a consequence of the influence of the Buchdahl parameter $\tilde{k}$. From the redshifted image of the lensed accretion disk, we observed that the Doppler blueshift can exceed the overall gravitational redshift, particularly in the left half-part of the disk. We showed that the Buchdahl-inspired spacetime significantly alters the shape and intensity of the accretion disk. The intensity of the lensed accretion disk is notably concentrated in a small region very close to the Buchdahl-inspired spacetime, indicated in red, approaching the outside observer. Interestingly, we found that the ray-traced images closely resemble the Schwarzschild black hole case. However, the size of the accretion disk region can be significantly influenced by the Buchdahl parameter $\tilde{k}$. 
In a wide literature on the accretion model, the well accepted model is based on Kerr hypothesis i.e., the accretor should be either slow rotating (even a Schwarzschild black hole) or a fast rotating Kerr black hole. However, there is no need to model a compact object in terms of a solution that is not one of the above two. We are motivated to propose an alternative to the Kerr hypothesis. In this connection, the special Buchdahl-inspired metric incorporates various physically interesting spacetime metrics including the traversable wormhole, a naked singularity or a Schwarzschild black hole, for different values of $\tilde{k}$. Our aim was to test this metric as an accretor model and have noticed that $\tilde{k}$ significantly effects the inner hollow region of the accretion disk. As a simulation, the model works quite well, however, we shall need a more thorough analysis in the future using observational data of accretion sources.

As a final remark, we like to mention that our analysis involved a vanishing cosmological constant.
Hence, it can be suspected that the final results which we obtained can be in tension with experimental data to explain QPOs. In general, within the precession model, there is a thorny issue related to the weird result that vacuum energy term seems to help fits to improve significance.
Accordingly, we report here that further analysis would be useful, including alternatives having cosmological constant contributions, since cosmological constant seems to be necessary.

\section{Acknowledgments}

This work is supported by the National Natural Science Foundation of China under Grants No. W2433018 and No. 11675143, and the National Key Research and Development Program of China under Grant No. 2020YFC2201503. We would like to thank anonymous reviewers for their important remarks and questions to improve this paper.


\end{document}